# Giant Magnetocaloric Effect in Re-entrant Ferromagnet PrMn$_{1.4}$Fe$_{0.6}$Ge$_2$

R. Zeng[1], J. L Wang[2,3], L. Lu[1], W.X. Li[1], S.J. Campbell[2], and S. X. Dou[1]

[1]Institute for Superconducting and Electronic Materials, University of Wollongong, NSW 2522, Australia

[2]School of Physical, Environmental and Mathematical Sciences, University of New South Wales, Australian Defense Force Academy, Canberra ACT 2600, Australia

[3]Bragg Institute, ANSTO, Lucas Heights, NSW 2234, Australia

Three first order magnetic phase transitions (FOMT) have been detected at $T_C^{Pr}$, $T_N^{inter}$ and $T_C^{inter}$ over the temperature range from 5 K to 340 K at fields up to 9 T in PrMn$_{1.4}$Fe$_{0.6}$Ge$_2$, and the magnetocaloric effect (MCE) around these transitions evaluated. The MCE of two FOMT from planar antiferromagnetism (AFl) to c-axis ferromagnetism (Fmc) around 168 K, and from the Fmc state to the c-axis AFmc state around 157 K have acceptable values compared with those of existing MCE systems. A giant magnetocaloric effect (GMCE) has been observed around 25.5 K associated with the field-induced FOMT from the AFmc to the Fmc+F(Pr) state with an additional Pr magnetic contribution. The MCE value $-\Delta S_M^{max}$ = 29.1 J kg$^{-1}$ K$^{-1}$ with field change $\Delta H$ = 7 T is comparable to and even larger than reported values for the best-performed MCE materials. In particular, the giant MCE value of 12.3 J kg$^{-1}$ K$^{-1}$ obtained for the relatively small field change from 0 to 1 T is very beneficial for applications, and this, together with the small magnetic and thermal hysteresis, suggests that PrMn$_{1.4}$Fe$_{0.6}$Ge$_2$ may be a promising candidate for magnetic refrigeration applications in the hydrogen liquefication temperature range.

Magnetic refrigeration based on the magnetocaloric effect is advantageous as it is environmentally friendly and offers an energy-efficient refrigeration mechanism. Cooling technology based on the magnetocaloric effect is considered extremely beneficial for the future in helping to address the two key issues of energy shortages and global warming. As such, it is highly desirable to explore new materials which offer prospects for a giant magnetocaloric effect (GMCE). The GMCE is generally associated with a first-order magnetic transition (FOMT) because of the large difference in magnetization between two adjacent magnetic phases. Thus, Pecharsky and Gschneidner discovered a GMCE in $Gd_5Ge_2Si_2$ originating from a first-order field induced structural and magnetic transition in this compound at 276 K [1] (see also [2,3]) with other GMCE ferromagnetic materials such as $MnFeP_{0.45}As_{0.55}$ [4], MnAs [5], and $La(FeSi)_{13}$ [6], subsequently having been discovered. These materials display a GMCE due to the first-order magnetic disorder-to-order transition from the paramagnetic to the ferromagnetic state near room temperature. There are also studies of paramagnetic salts, garnets, and molecular clusters [7-10] which can be utilized for low temperature cooling purposes.

Ternary rare-earth (R) compounds of the type $RT_2X_2$, where T = transition metal, X = Si, Ge, exhibit a large variety of structural and physical properties and have been studied extensively over the years. The interesting phenomenon of re-entrant ferromagnetism has been found in several of these $RT_2X_2$ compounds by controlling the interplay between the R–T and T–T exchange interactions through elemental substitution [11-13]. While there are several reports on MCE in $RT_2X_2$, [14-15], few reports [16] exist in the literature on an MCE associated with re-entrant ferromagnetic transitions. This may be due to the fact that most re-entrant phase transitions are order-to-order phase transitions between antiferromagnetic (AFM) and ferromagnetic (FM) states, and these kinds of transitions are not expected to show a very high MCE. The magnetic structures and properties of $PrMn_{2-x}Fe_xGe_2$ have been studied recently over the entire Mn and Fe concentration ranges e.g. [13]. In this Letter, interest is focussed on the magnetocaloric properties of the re-entrant ferromagnet $PrMn_{1.4}Fe_{0.6}Ge_2$. It has been found that a first-order temperature-induced and field-induced order (AFmc) to order (Fmc+F(Pr)) magnetic phase transition around 25.5 K can also induce giant MCE with low magnetic and thermal hysteresis. At the same time, a magneto-thermal switch phenomenon from positive (conventional) to negative (inverse) MCE has been observed in this re-entrant ferromagnet.

The polycrystalline $PrMn_{1.4}Fe_{0.6}Ge_2$ compound was prepared by arc melting the high purity elements on a water-cooled Cu hearth under purified argon gas. The mass loss of Mn during melting was compensated for by adding 2% excess Mn. The ingots were melted five times to attain homogeneity and then annealed at 900 °C for one week in an evacuated quartz tube. The room temperature x-ray diffraction pattern (CuKα radiation) confirms the single-phase nature of the $PrMn_{1.4}Fe_{0.6}Ge_2$ compound which crystallizes in the $ThCr_2Si_2$-type structure with no evidence of any impurity phases. The x-ray data were refined using a Rietveld profile fit (FULLPROF program) and the results of the refinements for the lattice parameters are $a = 0.4088(1)$ nm and $c = 1.0834(1)$ nm. The temperature dependence of the magnetization of $PrMn_{1.4}Fe_{0.6}Ge_2$ was measured in magnetic fields of 0.01 - 9 T in both cooling and heating modes over the temperature range 5 - 340 K with the magnetization also being measured from 20 - 60 K and 100 - 200 K for applied magnetic fields up to 7 T. The experiments were carried out on a physical properties measurement system (PPMS) and a superconducting quantum interference device (SQUID).

The temperature dependences of the $PrMn_{1.4}Fe_{0.6}Ge_2$ magnetization on cooling and heating from 5 to 340 K in magnetic fields from 0.01 T to 9 T are presented in Fig. 1(a). As shown in Fig 1(a), the temperature dependence of the magnetization for $PrMn_{1.4}Fe_{0.6}Ge_2$ reveals the presence of four magnetic transitions, similar to the case of $SmMn_2Ge_2$ [17–19]. Based on the magnetic structures of $PrMn_2Ge_2$ [20, 21] and the similarity in magnetic behaviour of $PrMn_{1.4}Fe_{0.6}Ge_2$ to that of $SmMn_2Ge_2$, it can be seen that with decreasing temperature from 340 K, PrMn$_{1.4}$Fe$_{0.6}$Ge$_2$ exhibits a transition from paramagnetism (PM) to an *ab*-plane antiferromagnetic state (AFl) [13] at Néel temperature, $T_N^{intra} \approx 333$ K. Around Curie temperature, $T_C^{inter} \approx 168$ K, this AFl structure gives way to a canted ferromagnetic structure Fmc-type. With further decrease in temperature, a canted antiferromagnetic structure (AFmc) state occurs around $T_N \approx 157$ K and, finally, there is a transition to an Fmc state plus an additional Pr-sublattice magnetic contribution (Fmc+F(Pr)) at $T_C^{Pr} \approx 25.5$K. Both PrMn$_{1.4}$Fe$_{0.6}$Ge$_2$ and SmMn$_2$Ge$_2$ exhibit similar magnetic behaviour[13] even though the values of the room temperature *a*-parameters differ by ~1 % (*a* = 0.4088 nm and *a* ~0.4045 nm [19] respectively). This common magnetic behaviour for disparate values of the *a*-parameter may be related to electronic effects as suggested in the case of NdMn$_{1.6}$Fe$_{0.4}$Ge$_2$ for which the ferromagnetic to antiferromagnetic transition takes place at a larger $d_{Mn-Mn}$ distance than in the pure NdMn$_2$Ge$_2$ compound [12].

In order to further understand these magnetic transitions, the thermomagnetic curves at different applied fields are presented in Fig. 1(b) as 3D surface plots. These curves indicate clearly that the width of the lower temperature Fmc phase region increases with applied magnetic field while the antiferromagnetic phase region decreases, thus demonstrating that the applied magnetic field destroys the AFmc state in PrMn$_{1.4}$Fe$_{0.6}$Ge$_2$ (this is also confirmed by the appearance of the field-induced metamagnetic phase transitions in the magnetization curves at 20 K and 60 K as shown in Fig. 2(a)). It is well known that in RMn$_2$Ge$_2$ compounds, even slight variations in the unit cell parameters due to external factors, such as pressure, field, temperature or chemical substitution, are sufficient to modify the interlayer Mn–Mn spacing, leading to magnetic phase transitions (e.g. ref. 16, 19). Such transitions are likely to be accompanied by an anomaly in the thermal expansion (magneto-volume effect) and can therefore be controlled by application of pressure or field. It has been reported that the width of the lower temperature antiferromagnetic phase in SmMn$_2$Ge$_2$ can be extended by applied hydrostatic pressure [16]. By comparison the present results demonstrate that an applied magnetic field on PrMn$_{1.4}$Fe$_{0.6}$Ge$_2$ has the opposite effect to that of applied pressure on SmMn$_2$Ge$_2$; this behaviour can be understood in terms of magnetostriction since magnetostriction effects generally induce tensile strain opposite to the pressure strain. It can also be seen from Fig. 1 (b) that with increasing field, the gorge-like AFmc region between $T_C^{Pr}$ and $T_N^{inter}$ becomes narrow and shallow and is almost closed above 9 T. The sharp cliff-like transition of AFmc to Fmc+F(Pr) at low field implies the unexpected giant magnetocaloric effect as outlined in the following analysis.

Fig. 2(a) shows the magnetization and demagnetization curves between 20 K and 60 K measured at 2 K intervals with the M-T-H 3D surface plots over the range 100 - 200 K (5 K intervals) shown in Fig. 2(b). It is clear that above $T_C^{Pr} \approx 25.5$ K and below $T_N^{Inter} \approx 157$ K, while the magnetization increases slowly with magnetic field in the low-field range and before increasing sharply at a critical field, the magnetization is not saturated at 7 T between 25.5 K and 157 K. The step in the magnetization curves indicates a field-induced AFM to FM or FM to AFM phase transition. In order to determine the transition type - first or second order - the Inoue–Shimizu model, which involves a Landau expansion of the magnetic free energy up to the sixth power of the total magnetization M, has been used:

$F(M,T) = (c1(T)/2)M^2 + (c2(T)/4)M^4 + (c3(T)/6) M^6 + \cdots - \mu_0 HM.$ (1)

It has been reported [24] that the order of a magnetic transition is related to the sign of the Landau coefficient $c2(T)$. A transition is expected to be first order when $c2(T_C)$ is negative, whereas it will be second-order for a positive $c2(T_C)$. The sign of $c2(T_C)$ can be determined by means of Arrott plots - if the Arrott plot is S-shaped near $T_c$, then $c2(T_C)$ is negative, otherwise, $c2(T_C)$ is positive. The Arrott plots of H/M versus $M^2$ at 2 K intervals between 20 K and 60 K and at selected temperatures between 120 K and 200 K are shown in Figs 2(c) and (d), respectively. It can be seen that the magnetic transitions from AFl to Fmc above 168K, from Fmc to AFmc below 157 K, and from AFmc to Fmc+F(Pr) below 25.5 K are first order transitions in applied magnetic field.

The thermal hysteresis (~ 4.5 K) and magnetic hysteresis (~ 0.2 T) near the AFmc to Fmc+F(Pr) transition around 25 K (as determined from Figs. 1(a) and 2(a) respectively), are relatively small compared to typical values for materials which exhibit a GMCE. For example, near the magnetic transition temperature in $Gd_5(Ge_{1-x}Si_x)_4$ [2] the thermal hysteresis is about 7.4 K and the magnetic hysteresis is about 1 T. This small magnetic hysteretic effect is believed to originate from the field-induced FM state at low temperatures [22]. Fig. 3 shows the magnetic phase diagram (magnetic field versus temperature) derived for $PrMn_{1.4}Fe_{0.6}Ge_2$ from the above analysis (the phase boundaries in Fig. 3 were determined in the standard way from the extreme values and turning points of graphs of dM/dH versus H, dM/dT versus T and 1/M versus T; see for example [13]). The phase diagram for $PrMn_{1.4}Fe_{0.6}Ge_2$ in Fig. 3 enables us to understand the unique MCE behaviour of this compound.

The isothermal entropy change, corresponding to a magnetic field change $\Delta H$ starting from zero field, was derived from the magnetization data by means of the following expression which can be obtained from the Maxwell relation:

$$\Delta S_M (T,H) = \int_0^H (\delta M/\delta T)_H \delta H \qquad (2)$$

The temperature dependence of the isothermal magnetic entropy change $-\Delta S_M (T, H)$ calculated from Eq. (2) for different magnetic-field changes around the three first order magnetic phase transition temperatures - 25.5 K, 157 K, and 168 K - are presented in Figs. 4(a) and (b). The values of $-\Delta S_M^{max}$ at 25.5 K are 12.3, 18.5, 22.4, 24.7, 26.6, 28.2, and 29.1 J kg$^{-1}$ K$^{-1}$ for magnetic-field changes from 0 to 1, 2, 3, 4, 5, 6, and 7 T, respectively. These values are larger than those reported for Gd (5.1 J kg$^{-1}$ K$^{-1}$ for 2 T) or $Gd_5(Si_2Ge_2)$ (-14.1 J kg$^{-1}$ K$^{-1}$ for 2 T) [2]. The giant MCE value $-\Delta S_M^{max}$ = 12.3 J kg$^{-1}$ K$^{-1}$ obtained for the relatively small field change from 0 to 1 T, is very beneficial for applications. The maxima of $-\Delta S_M{}^{max}$ at 25.5 K for the different magnetic-field changes correspond to the temperature and field induced first order AFmc to Fmc phase transition.

The curves of $-\Delta S_M$ versus T shown in Fig. 4(b), provide valuable information about the nature of the two other magnetic ordering transitions in PrMn$_{1.4}$Fe$_{0.6}$Ge$_2$. Around $T_N{}^{inter} \approx 157$ K, $-\Delta S_M$ is negative (inverse MCE) corresponding to the magnetic transition from the Fmc to the AFmc state (Fig. 3), but it changes to positive values with increasing temperature around $T_C \approx 168$ K, corresponding to the magnetic transition from the AFl to the Fmc state (Fig. 3). An inverse MCE has been often observed in systems displaying first-order magnetic transitions and the origin is the same for all of them. Due to the presence of mixed exchange interactions, the applied magnetic field leads to a further spin-disordered state near the transition temperature, increasing the configurational entropy [8]. As shown in Fig. 4(b), the $-\Delta S_M{}^{max}$ peak position moves to the low temperature side with increasing magnetic field. This behavior, which is commonly observed in field-induced first order phase transition MCE materials [23], is due to the large thermal hysteresis (Fig. 1 (a)) and the large shift of critical field to high field with decreasing temperature around $T_N$ for the Fmc to AFmc transition (Fig.4(c)).

To summarize, both conventional and inverse magnetocaloric effects have been observed in polycrystalline PrMn$_{1.4}$Fe$_{0.6}$Ge$_2$. The unexpected exhibition of a giant magnetocaloric effect around 25.5 K is associated with the field-induced FOMT from the AFmc to the Fmc+ F(Pr) state. The giant MCE values are comparable to and even larger than those reported for the best-known MCE materials. In particular, the giant MCE value $-\Delta S_M{}^{max} = 12.3$ J kg$^{-1}$ K$^{-1}$, obtained for a small field change from 0 to 1 T, and, together with the small hysteresis, is very beneficial for applications. This suggests that PrMn$_{1.4}$Fe$_{0.6}$Ge$_2$ may be a promising candidate to be applied in magnetic refrigeration in the low temperature hydrogen liquefication range. Our observations indicate that systems which exhibit re-entrant magnetism also offer a route for giant MCE. In addition, this study provides significant opportunities for exploring new GMCE materials, such as the present layered $ThCr_2Si_2$-type structure, by controlling magnetic transitions in re-entrant magnetism through varying external factors, such as pressure, field, temperature, or chemical substitution, to vary the unit-cell parameters and to modify the interlayer Mn–Mn spacing.

The authors thank Dr. T. Silver for her help and useful discussions. This work is supported by the Australian Research Council through a Discovery project (project ID: DP0879070).

**Figure Captions**

Fig. 1. (a) Temperature dependence of the magnetization for PrMn$_{1.4}$Fe$_{0.6}$Ge$_2$ on cooling (open symbol) and heating (full symbol) over the range 5 to 340K for magnetic fields from 0.01 T to 9T, and (b) the M-H-T 3D surface plots (cooling curves; for convenience the transition temperatures in zero field are indicated by arrows, cf. Fig 1(a)).

Fig. 2. (a) Magnetization and demagnetization curves; (b) Arrott plots H/M versus M$^2$ at 2 K intervals between 20 and 60 K; (c) M-T-H 3D surface plots at 5 K intervals between 100 and 200 K and (d) Arrott plots H/M versus M$^2$ at the temperatures indicated.

Fig. 3. Magnetic phase diagram of PrMn$_{1.4}$Fe$_{0.6}$Ge$_2$ for applied magnetic fields in the range 0 - 9 T and temperatures from 5 - 340 K. The phase boundaries were determined from dM/dT versus T and 1/M versus T curves as discussed in the text. The thermal hysteresis around three transitions determined from the cooling and heating magnetic phase diagrams are $\Delta T_C^{Pr}$ ~4.5 K, $\Delta T_N^{Inter}$ ~7K to 16 K and $\Delta T_C^{inter}$ ~24 K, respectively.

Fig. 4. Temperature dependence of the isothermal magnetic entropy change $-\Delta S_M$ (T, H) around three 1$^{st}$ order magnetic phase transition temperatures: (a) T$_c^{Pr}$; and (b) T$_N^{inter}$ and T$_C^{inter}$. (c) dM/dH versus H curves at T = 115 K to 155 K at 5 K intervals. The insert shows the temperature dependence of the critical magnetic field (H$_{cr}$) of the metamagnetic transition.

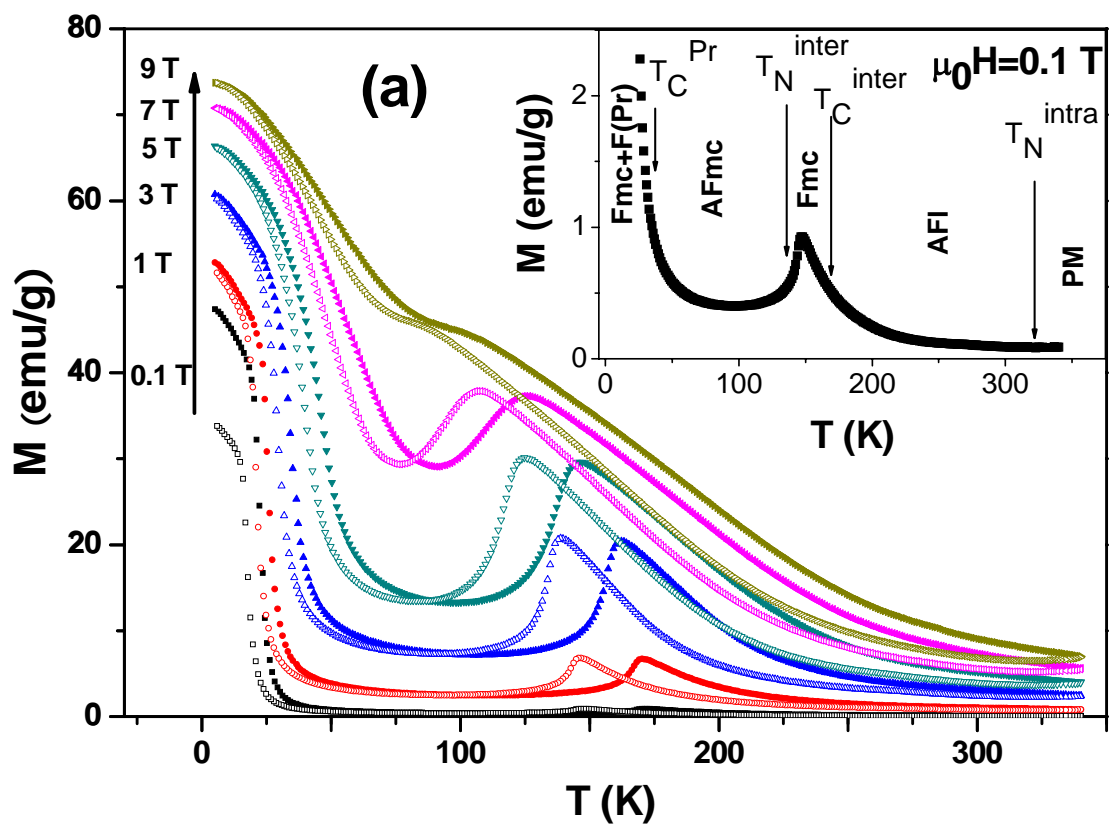

Fig. 1 (a).

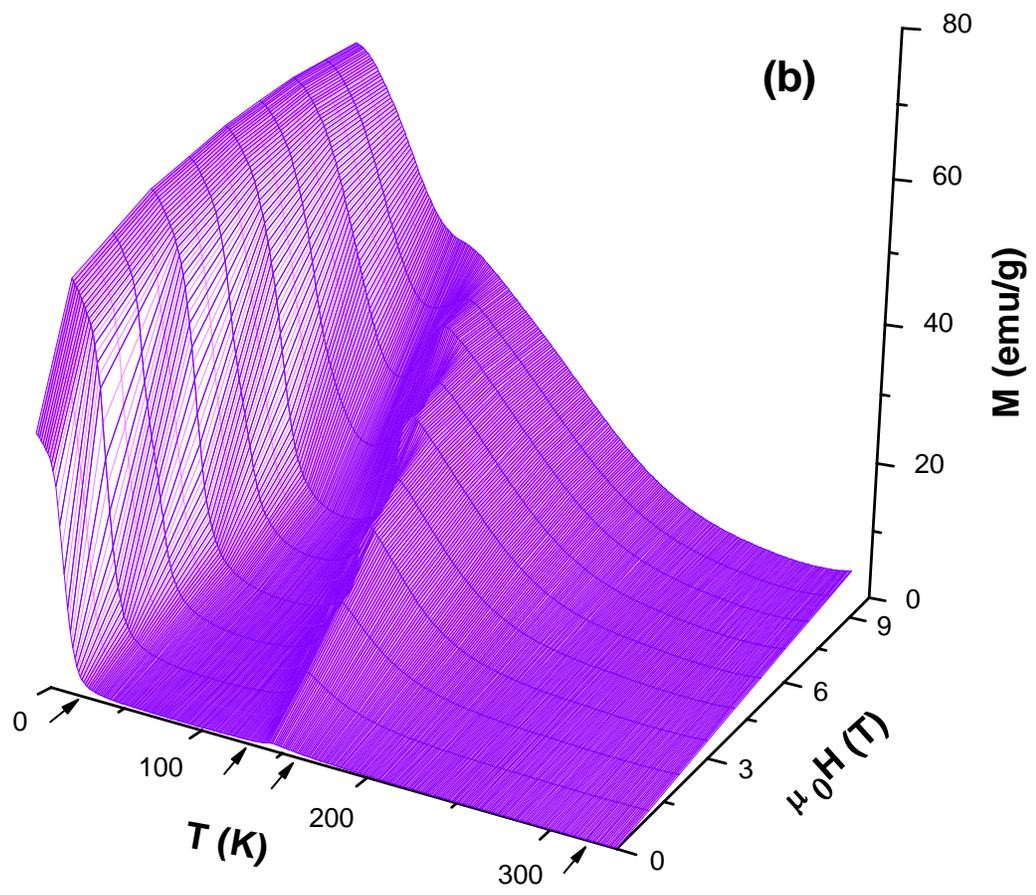

Fig.1 (b).

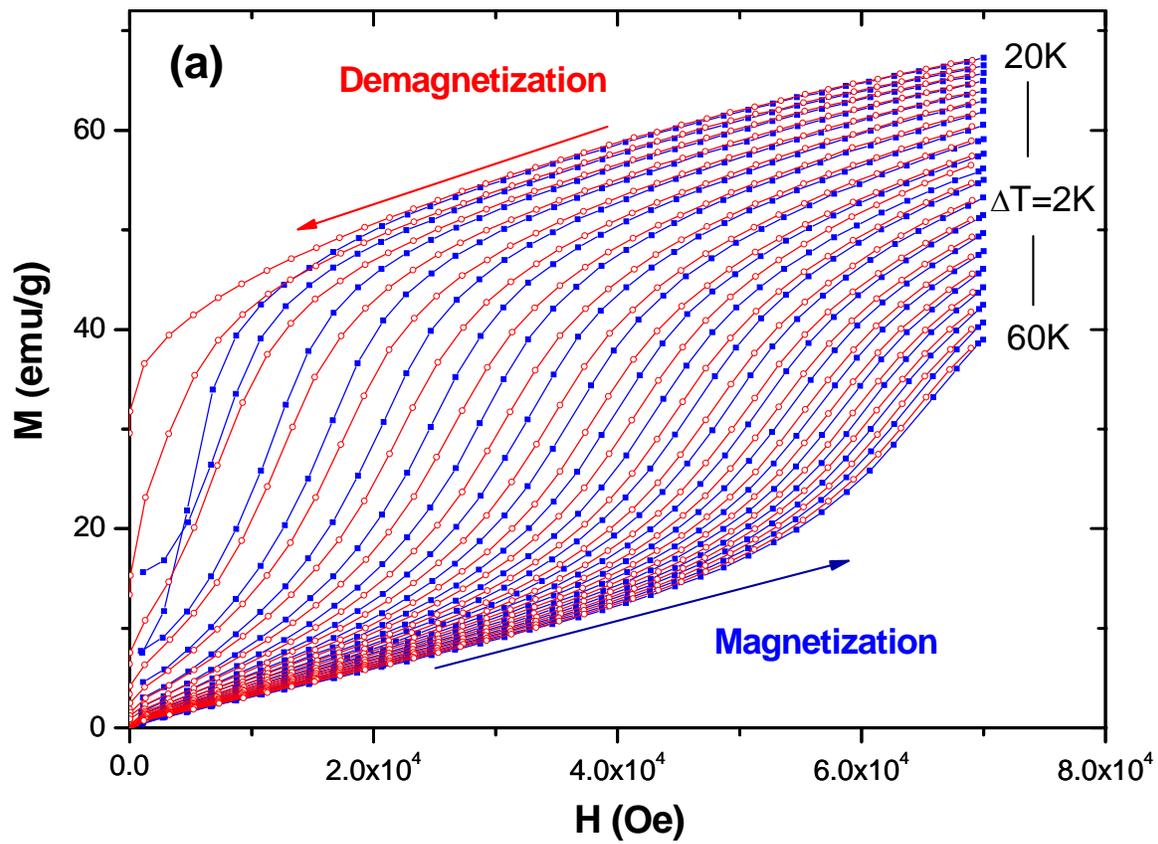

Fig. 2 (a)

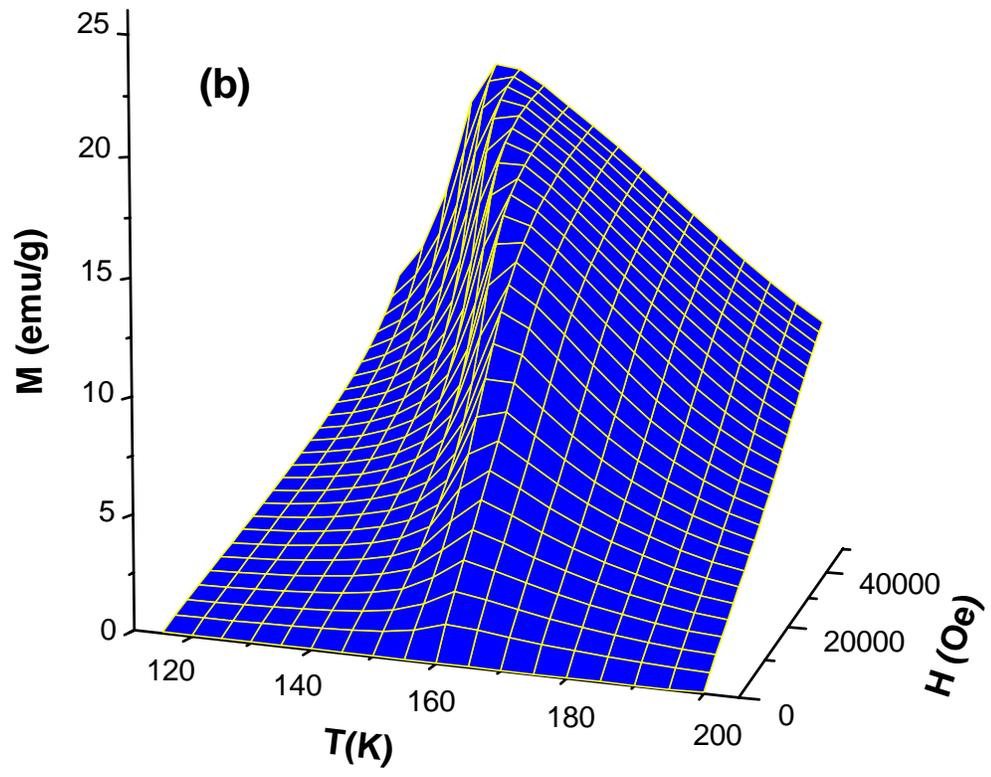

Fig. 2 (b).

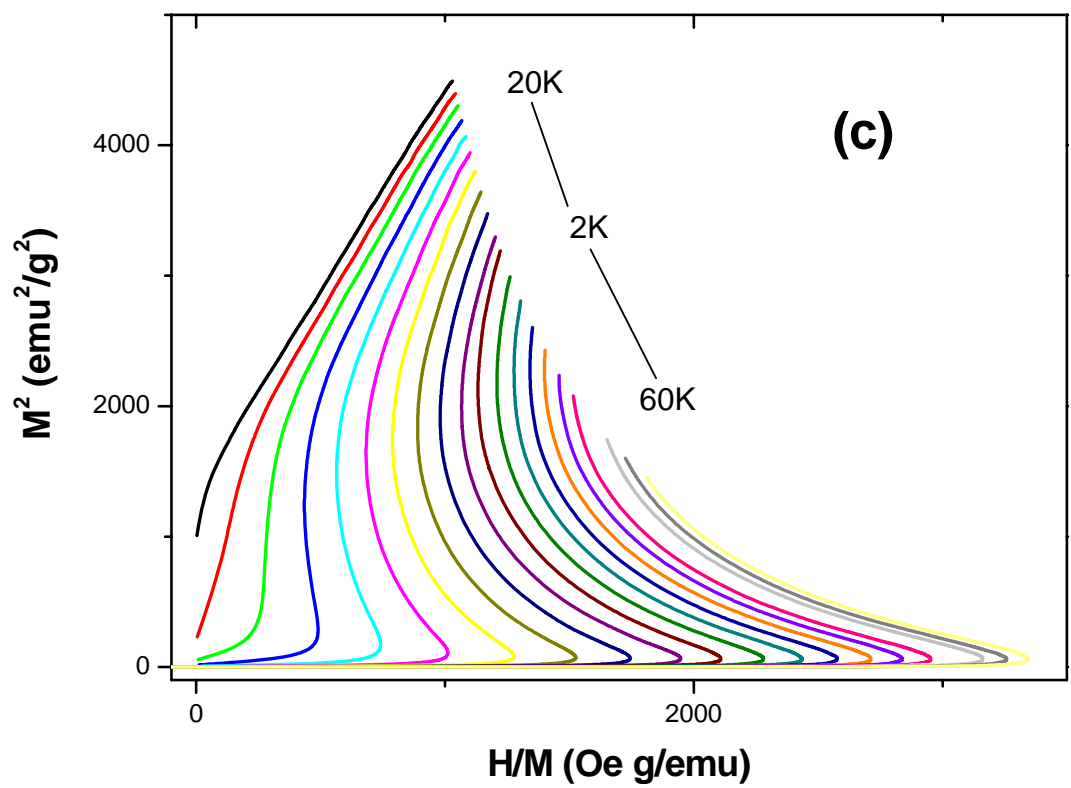

Fig. 2 (c).

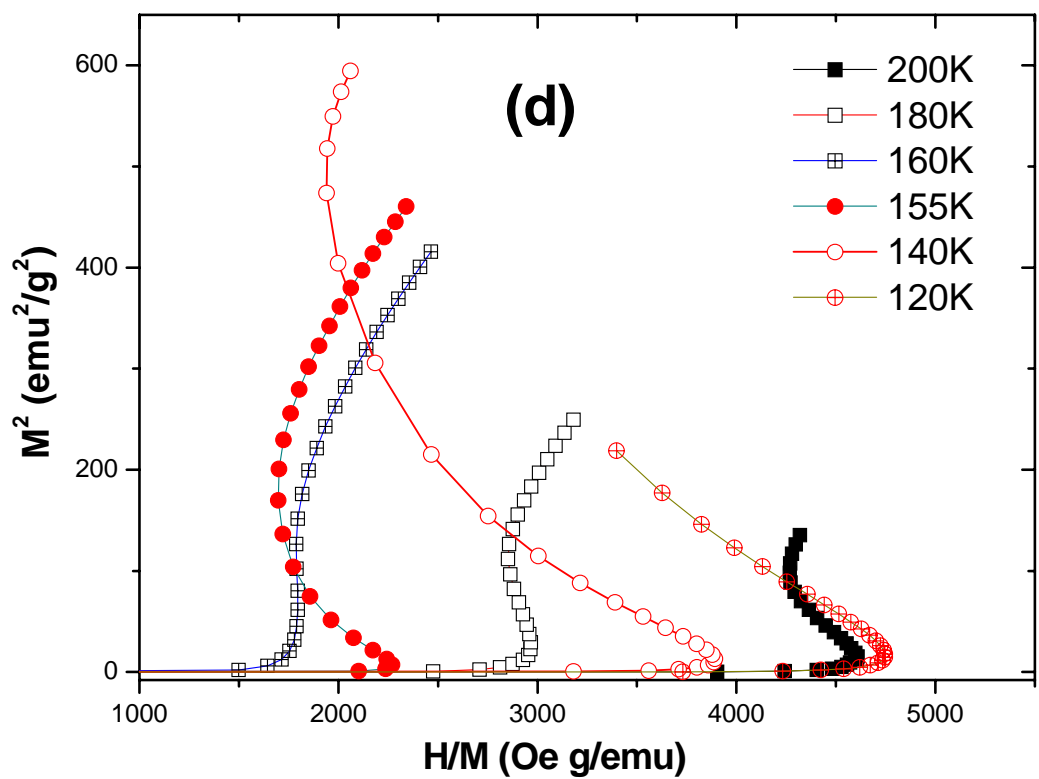

Fig. 2 (d).

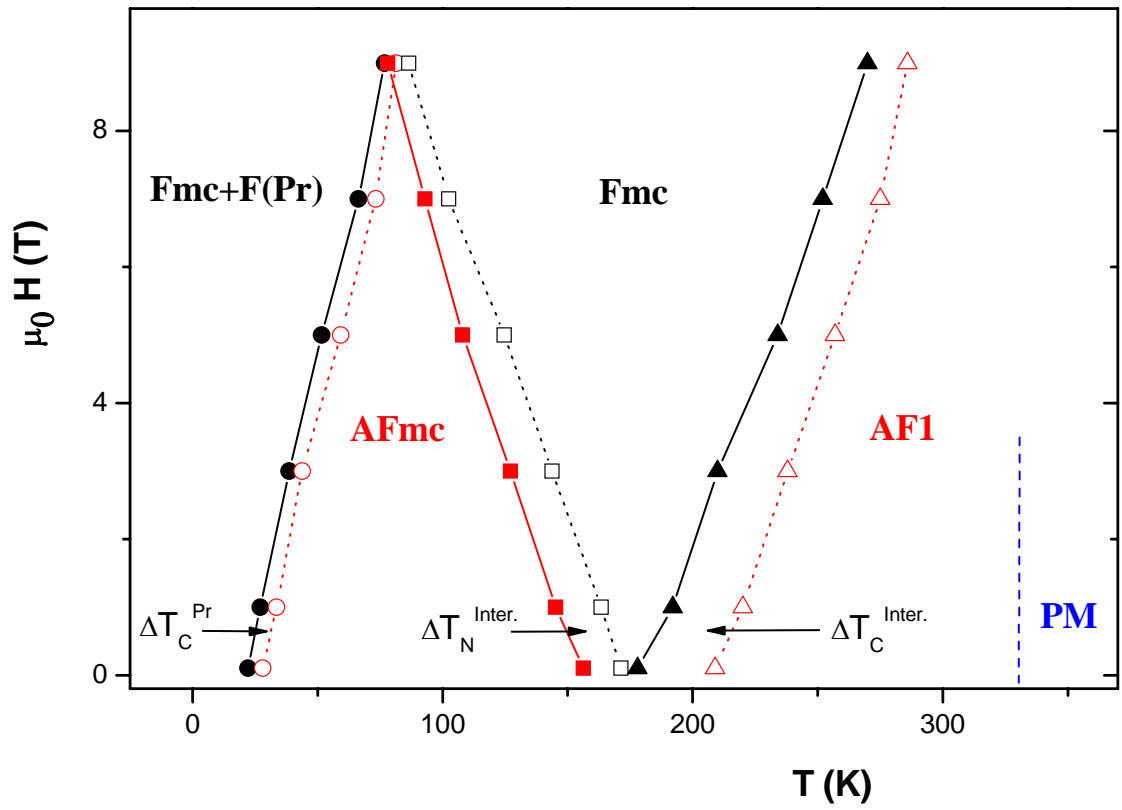

Fig. 3.

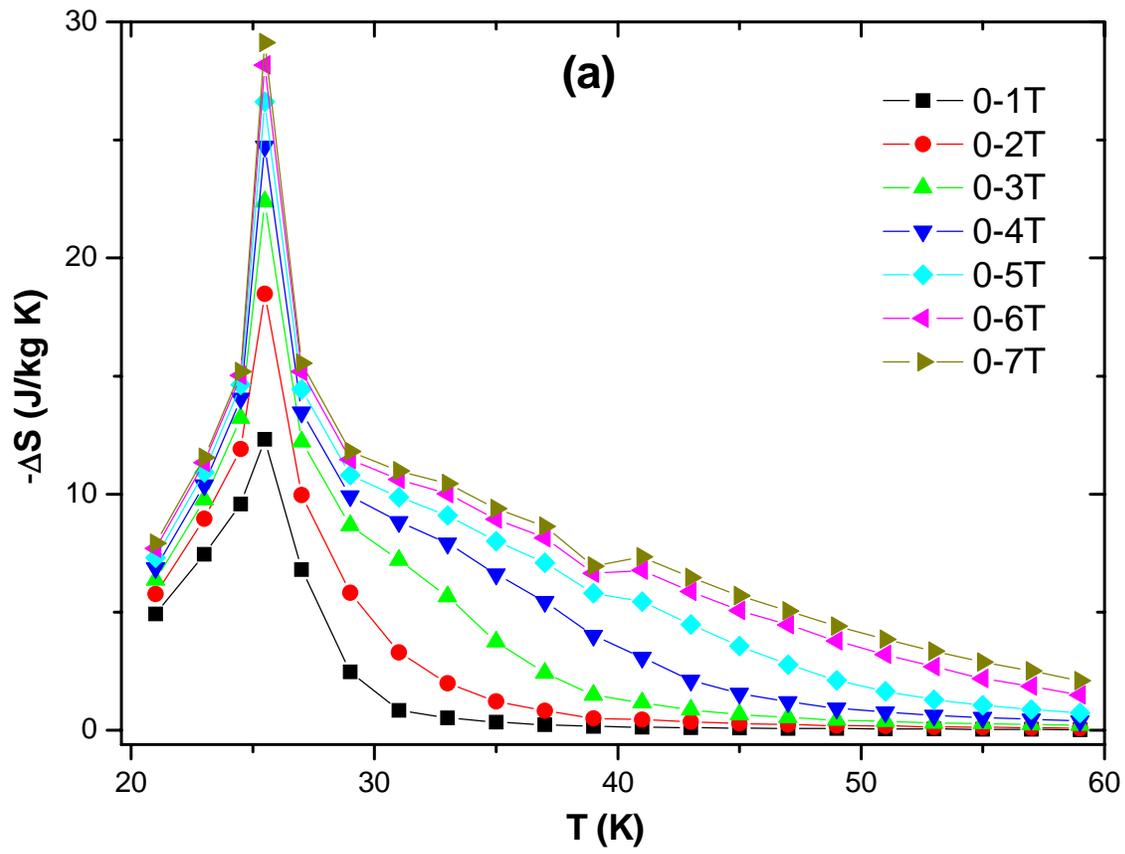

Fig. 4 (a).

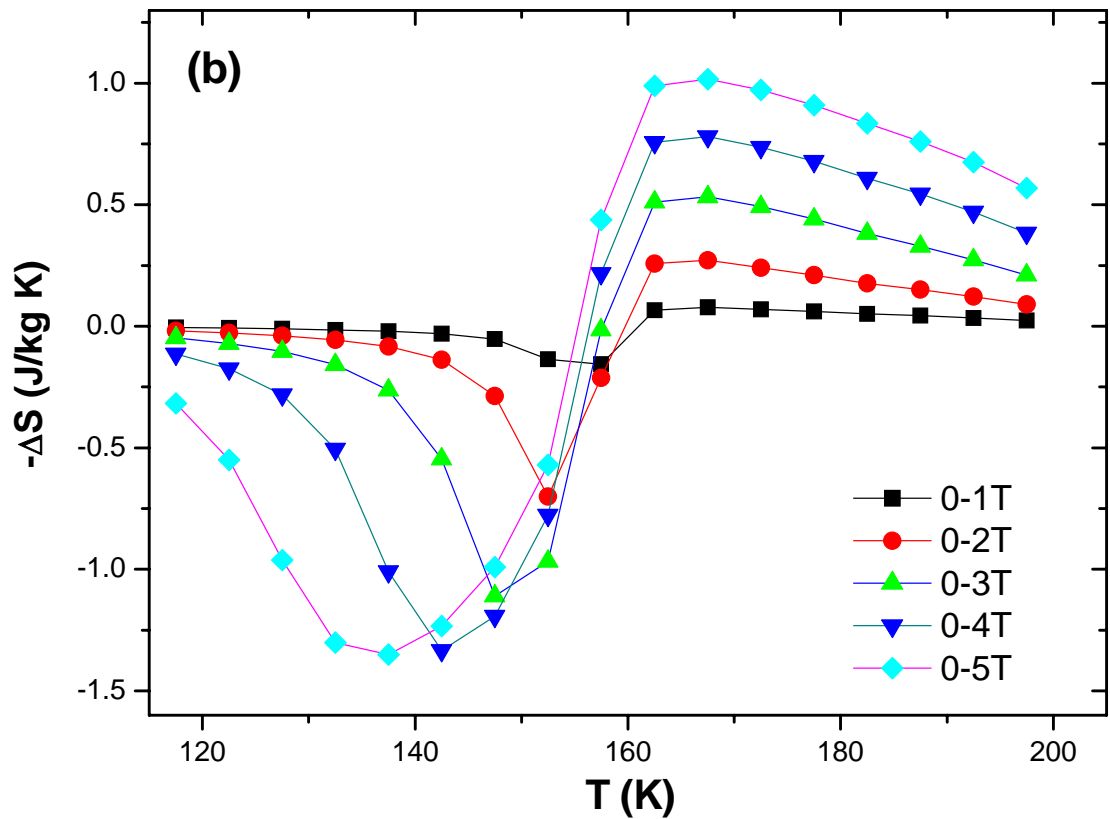

Fig. 4 (b).

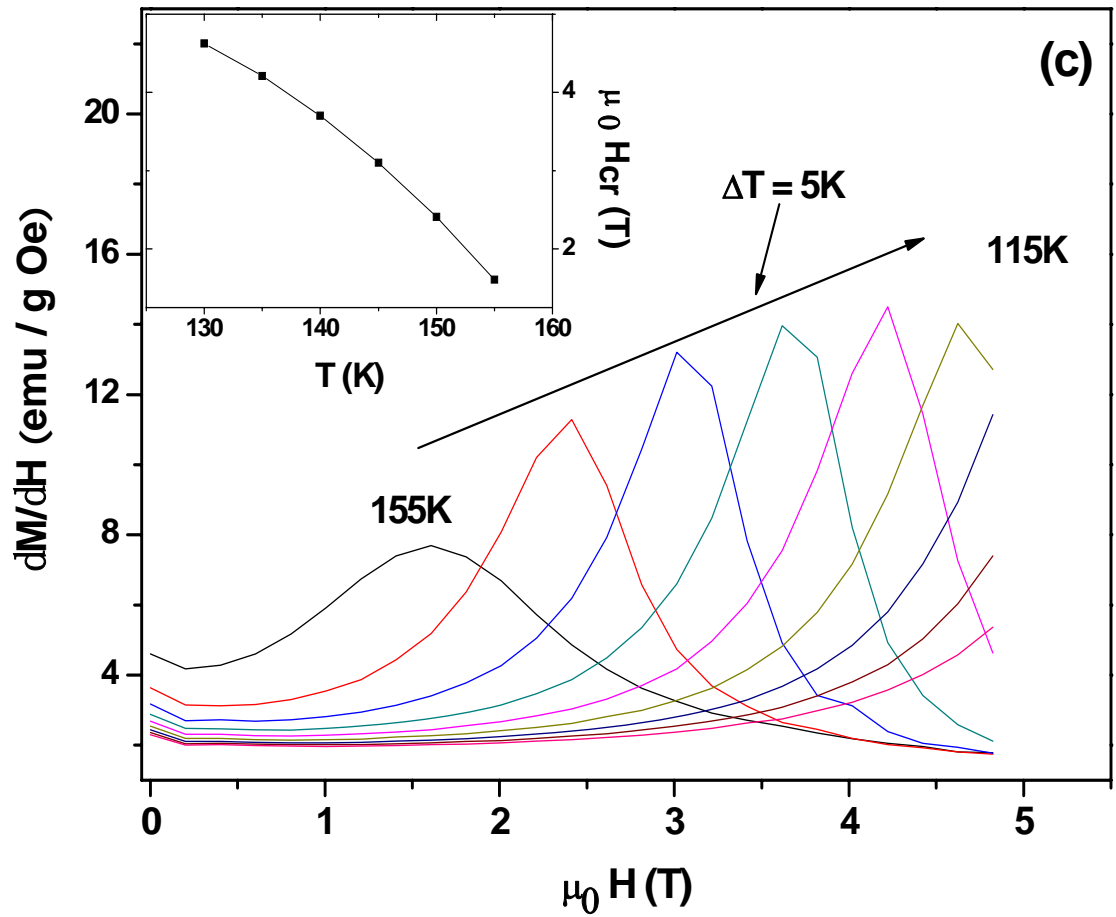

Fig. 4(c).